# ¿Vapor Pressure in a Paramagnetic Solid?

**Manuel Malaver**

Bijective Physics Institute, Bijective Physics Group, Idrija, Slovenia
Maritime University of the Caribbean, Catia la Mar, Venezuela.

Email Address:  mmf.umc@gmail.com

**Abstract:** In this paper, we obtain an analytical expression for the vapor pressure of a paramagnetic solid for high temperatures. We have considered the behavior of magnetic materials in the presence of an external magnetic field using the thermodynamical analysis and the elements of statistical mechanics in microscopic systems. We found that the vapor pressure depends on the magnetic susceptibility of material and the external field applied.

## 1. Introduction

   The magnetism is a type of associated physical phenomena with the orbital and spin motions of electrons and the interaction of electrons with each other [1]. The electric currents and the atomic magnetic moment of elementary particles generate magnetic fields that can act on other currents and magnetic materials. The most common effect occurs in ferromagnetic solids which is possible when atoms are rearrange in such a way that magnetic atomic moments can interact to align parallel to each other [2,3]. In quantum mechanics, the ferromagnetism can be described as a parallel alignment of magnetic moments what depends of the interaction between neighbouring moments [4].
    Although ferromagnetism is the cause of many frequently observed magnetism effects, not all materials respond equally to magnetic fields because the interaction between magnetic atomic moments is very weak while in other materials this interaction is stronger [1]. In the paramagnetic materials there are unpaired electrons which freely align their magnetic moment in any direction and when an external field is applied these magnetic moments align in the same direction of applied field making it more intense [2].
    Diamagnetism is an intrinsic property of all materials and is the tendency of materials to be repelled by an applied magnetic field because of the presence no unpaired electrons so that the atomic magnetic moments not produce any effect [5]. The other forms of magnetism as paramagnetism or ferromagnetism are much stronger in a material and the diamagnetic contribution is very negligible [1].
   In the statistical and microscopic description of any system, the partition function plays determinant role and is defined as the total sum of states of the system [6]:

$$Z = \sum g(E_n) e^{-\frac{E_n}{kT}} \qquad (1)$$

where *n* labels the total energy $E_n$ , *k* is the Boltzman´s constant and $g(E_n)$ it is the number of degenerate states with the same energy $E_n$. The use of the equation (1) allows to obtain a statistical and microscopic description of the energy and the entropy [6,7].

The concepts of the statistical mechanics must be considered if we want to do a microscopic description of a physical system. Recently Mäkelä [8,9] constructed a microscopic model of "Stretched Horizon " of a Schwarzschild and Reissner-Nordström black holes and obtained an analytical expression for the partition function from the point of view of an observer on its stretched horizon. Malaver [10,11,12] studied the behavior of the thermal capacity $C_V$ for Schwarzschild and Reissner-Nordström black holes when $T>>T_C$ and $T<<T_C$ where $T_C$ is the characteristic temperature and found that the value for $C_V$ if $T>>T_C$ is the same that would be obtained in an ideal diatomic gas if are considered the rotational and translational degrees of freedom, respectively. Viaggiu [13] present a statistical analysis in gravitons and derived equations for the partition function and the mean energy. Malaver [14] obtained an analytical expression for the thermal capacity for gravitons and study the behavior of $C_V$ in the limit of high and low temperature, Mandl [7] used the concepts of statistical thermodynamics to understand the behavior of paramagnetic solids and García-Colín Scherer [15] presents a thermodynamic analysis in magnetic systems. Also Zemansky and Dittman [16] apply statistical mechanics in order to study different types of magnetic materials.

In this paper we have deduced an analytical expression for the vapor pressure in a paramagnetic solid. We found that the vapor pressure will depend on the magnetic susceptibility of material and the external magnetic field applied. This work is outlined in the following manner: the section II the behavior of paramagnetic solids are studied; in section III, we present the statistical thermodynamics of magnetic materials; in section IV is obtained an expression for the vapor pressure for a paramagnetic solid; in Section V, presents the conclusions of this study.

## 2. Behavior of Paramagnetic Solids

In a paramagnetic material, the atoms contain permanent magnetic moments and the induced magnetic fields are aligned in the direction of the applied magnetic field [3]. Paramagnetism is due to the presence of unpaired electrons in the materials and the atoms with semifilled atomic orbitals are paramagnetic [1]. Examples of paramagnetic materials are aluminium, oxygen and titanium. The tendency of the magnetic moments to align with the direction of the field is counteracted by the thermal movement that disorders the magnetic dipoles. In the magnetic materials, the magnetization M is proportional to intensity of magnetic field H and the mathematical expression is [2]

$$M = \chi H \qquad (2)$$

where is χ the magnetic susceptibility of material and depends on the temperature $T$ as follow

$$\chi = \frac{C}{T} \qquad (3)$$

The eq. (3) is known as Curie´s law and C is a positive constant characteristic of material.

The paramagnetic susceptibility can be calculated for a type of system that contains separate atoms where there are unpaired electrons in an external magnetic field **B**. Following Eisberg-Resnick [2] if $n$ is the number of unpaired electrons per unit volume, $n_-$ is the volumetric density of moments that are parallel to the field and $n_+$ represents the density for the moments that are antiparallel, then $n = n_- + n_+$ . For a parallel alignment of the magnetic moment $\mu$ the magnetic potential energy is -$\mu B$ and for an antiparallel alignment the energy is $\mu B$. From the Boltzman´s distribution for each state of energy

$$n_- = n e^{\mu B/kT} \qquad \text{and} \qquad n_+ = n e^{-\mu B/kT} \qquad (4)$$

where $k$ is the Boltzman´s constant. The resultant magnetization is given by

$$M = \mu(n_- - n_+) = \mu n (e^{\mu B/kT} - e^{-\mu B/kT}) \qquad (5)$$

and the average magnetic moment can be written as

$$\bar{\mu} = \frac{M}{n} = \mu \frac{e^{\mu B/kT} - e^{-\mu B/kT}}{e^{\mu B/kT} + e^{\mu B/kT}} \qquad (6)$$

As the limiting case, let us consider the situation at very high temperature and weak field, then when $\mu B << kT$ and with the development of exponential terms we obtain

$$\bar{\mu} = \frac{M}{n} = \mu \frac{(1+\mu B/kT) - (1-\mu B/kT)}{(1+\mu B/kT) + (1-\mu B/kT)} = \frac{\mu^2 B}{kT} \qquad (7)$$

and the paramagnetic susceptibility is given for

$$\chi = \frac{M}{H} = \frac{n\bar{\mu}}{H} \approx \frac{n\mu^2 B}{kTH} \qquad (8)$$

But the magnetization also can be written as [2]

$$M = \frac{\chi B}{\mu_0 (1+\chi)} \qquad (9)$$

where $\mu_0$ is the vacuum permeability. Then when $\chi \ll 1$ and with eq. (2) we have

$$B \approx \mu_0 H \qquad (10)$$

Substituting (10) in (8) we can obtain

$$\chi \approx \frac{\mu_0 n \mu^2}{kT} \qquad (11)$$

The eq.(11) is an approximation of Curie's law with $C = \mu_0 n \mu^2 / k$ and the paramagnetic susceptibility varies inversely with the temperature. According to (7) the alignment of the atomic magnetic moments depends on the presence of field and if $B=0$ the thermal motion orients the moments randomly and the net magnetization is null [2,7].

## 3. Statistical Thermodynamics of Paramagnetic Materials

Following Mandl [7] we have that in separate atoms with unpaired electrons y that can exist in two states with energy $\pm \mu B$, the partition function for solid of N dipoles

$$Z = \left( e^{\frac{\mu B}{kT}} + e^{-\frac{\mu B}{kT}} \right)^N \qquad (12)$$

Defining the average statistical energy with the following expression [4]

$$\overline{E} = -\frac{\partial \ln Z}{\partial \beta} \qquad (13)$$

With the eq. (12), for the magnetic energy of N dipoles we obtain

$$E = N\mu B \frac{\left(e^{-\frac{\mu B}{kT}} - e^{\frac{\mu B}{kT}}\right)}{\left(e^{\frac{\mu B}{kT}} + e^{-\frac{\mu B}{kT}}\right)} \tag{14}$$

For the determination of the entropy in the paramagnetic solid can be use the expression for the Helmholtz free energy $A = E\text{-}TS$. In statistical mechanics, the work function o Helmholtz free energy A is defined as

$$A = -kT \ln Z \tag{15}$$

Substituting eq. (12) in (15) we have for the work function

$$A = -NkT \ln \left(e^{\frac{\mu B}{kT}} + e^{-\frac{\mu B}{kT}}\right) \tag{16}$$

With (14), (16) and the expression for A, the entropy of the system can be written as

$$S = Nk \left[ \frac{\left(e^{-\frac{\mu B}{kT}} - e^{\frac{\mu B}{kT}}\right)}{\left(e^{\frac{\mu B}{kT}} + e^{-\frac{\mu B}{kT}}\right)} + \ln \left(e^{\frac{\mu B}{kT}} + e^{-\frac{\mu B}{kT}}\right) \right] \tag{17}$$

In the limit of high temperature and weak field $\mu B << kT$ and with the development of exponential terms, we have for (17)

$$S = Nk \left[ \frac{1 - \frac{\mu B}{kT} - \left(1 + \frac{\mu B}{kT}\right)}{\left(1 + \frac{\mu B}{kT} + 1 - \frac{\mu B}{kT}\right)} + \ln \left(1 + \frac{\mu B}{kT} + 1 - \frac{\mu B}{kT}\right) \right]$$

$$\tag{18}$$

For these approximations we obtain for the entropy of system

$$S = Nk \ln 2 \qquad (19)$$

The eq.(19) implies that for N dipoles all states are equally possible and the orientation of atomic magnetic moments is disordered.

## 4. Determination of the Vapor Pressure in a Paramagnetic Solid

Based on Einstein's model, consider now that a paramagnetic material behaves as a crystalline solid and assume that the atoms of solid behave as oscillators that vibrate at same frequency $v$. If the vibrations are harmonic then the energy of harmonic oscillator is given by

$$\varepsilon_v = \left(v + \frac{1}{2}\right)hv \qquad (20)$$

According Nash [6] the eq.(20) then permits us to express in terms of $v$ and Planck´s constant $h$, the energy $\varepsilon_v$ associated with any one of the permissible quantum states corresponding to integral values of the quantum number $v$. If as usual we identify our reference zero of energy with the ground-state energy $\varepsilon_0=0$, we obtain $\varepsilon_1=hv$, $\varepsilon_2=2hv$ and so on. We can then easily formulate the partition function for a harmonic oscillator in terms of the empirically determinable value for $v$ which characterizes that species of oscillator

$$z = 1 + e^{-\frac{hv}{kT}} + e^{-\frac{2hv}{kT}} + e^{-\frac{3hv}{kT}} + \ldots \qquad (21)$$

This expression has the form

$$z = 1 + e^{-x} + e^{-2x} + e^{-3x} + \ldots \qquad (22)$$

and this series is equal to the function [3,4]

$$f(x) = \frac{1}{1-e^{-x}} \tag{23}$$

Then with the harmonic oscillator we can express the partition function in a closed analytic form

$$z = \frac{1}{1-e^{-\frac{h\nu}{kT}}} \tag{24}$$

Given this simple expression for the partition function, with eq. (13) we can calculate the energy $E$ of $N$ harmonic oscillators of solid

$$E = NkT^2 \left[\frac{d \ln z}{dT}\right]_V = NkT^2 \frac{d}{dT} \ln \frac{1}{1-e^{-\frac{h\nu}{kT}}} \tag{25}$$

and is obtained that

$$E = Nk\left(\frac{h\nu}{k}\right) \frac{1}{e^{\frac{h\nu}{kT}} - 1} \tag{26}$$

From the eq.(26) we have the following expression for the energy of $3N$ harmonic oscillators of crystalline solid

$$E = 3Nk\left(\frac{h\nu}{k}\right) \frac{1}{e^{\frac{h\nu}{kT}} - 1} \tag{27}$$

With (15), (24), (27) and from $A = E-TS$ can be determined the entropy of the vibrations of crystalline solid [6,7]

$$S = 3Nk\left[\frac{h\nu}{kT} \frac{1}{e^{\frac{h\nu}{kT}} - 1} + \ln\left(\frac{1}{1-e^{-\frac{h\nu}{kT}}}\right)\right] \tag{28}$$

The eq.(28) is rewritten in the following manner

$$S = 3Nk\left[\frac{h\nu}{kT}\frac{1}{e^{\frac{h\nu}{kT}}-1} - \ln\left(1-e^{-\frac{h\nu}{kT}}\right)\right] \tag{29}$$

We now define a parameter $\theta = h\nu/k$ where $\theta$ has dimensions of temperature and (29) can be express in a form more compact

$$S = 3Nk\left[\frac{\theta}{T}\frac{1}{e^{\frac{\theta}{T}}-1} - \ln\left(1-e^{-\frac{\theta}{T}}\right)\right] \tag{30}$$

If $N$ is Avogadro´s number then $Nk = R$ where $R$ is the ideal gas constant and we obtain from eq.(30) for the molar entropy of the solid

$$S = 3R\left[\frac{\theta}{T}\frac{1}{e^{\frac{\theta}{T}}-1} - \ln\left(1-e^{-\frac{\theta}{T}}\right)\right] \tag{31}$$

Let us consider the situation at very high temperature when $T \gg \theta$. As the function $e^x$ is equivalent to the sum of the infinite series $1 + x + x^2/2! + x^3/3! + \ldots$ can be developed the exponential terms in (8) and we find [7]

$$S = R[3\ln T - 3\ln \theta + 3] \tag{32}$$

Total entropy for the solid paramagnetic for high temperatures will be given by the vibrational contribution (32) and the paramagnetic contribution (19) that is

$$S = R[3\ln T - 3\ln \theta + 3 + \ln 2] \tag{33}$$

For the determination of the solid's vapor pressure, it must be assumed that the vapor behaves as an ideal monoatomic gas and its partition function will be given by [6]

$$Z = \frac{1}{N!}\left[\left(\frac{2\pi mkT}{h^2}\right)^{\frac{3}{2}} V\right]^N \tag{34}$$

Where $m$ is the mass of the gas particles and $V$ is the volume of the container containing it. Substituting (34) in equation (15), we express the Helmholtz free energy of an ideal monoatomic gas as follows

$$A = -NkT\left\{1 + \ln\left[\left(\frac{2\pi mkT}{h^2}\right)^{\frac{3}{2}} \frac{V}{N}\right]\right\} \tag{35}$$

For the thermodynamic relation

$$dA = -PdV - SdT \tag{36}$$

When the temperature is constant, (36) is reduced to

$$P = -\left[\frac{dA}{dV}\right]_T \tag{37}$$

then we have now for the pressure $P$

$$P = NkT\frac{d}{dV}\left\{\ln V + \ln\frac{1}{N}\left[\frac{2\pi mkT}{h^2}\right]^{\frac{3}{2}}\right\}_T \tag{38}$$

and we obtain for an ideal monoatomic gas

$$P = NkT\frac{1}{V} \qquad (39)$$

or for one mole of gas $PV=RT$ and with the equation (13) we can now express the energy of an ideal gas as

$$E = kT^2\left[\frac{d\ln Z}{dT}\right]_V = kT^2 \frac{d}{dT}\left\{\ln\frac{1}{N!}\left(\left[\frac{2\pi mkT}{h^2}\right]^{\frac{3}{2}}V\right)^N\right\}_V \qquad (40)$$

and the equation (40) can be written as follows

$$E = kT^2 \frac{d}{dT}\ln T^{\frac{3N}{2}} = \frac{3}{2}NkT \qquad (41)$$

For a paramagnetic gas the Helmholtz free energy $A$ is given by [16]

$$A = E - TS - \mu_0 MH \qquad (42)$$

Substituting (35) and (41) in eq. (42) we find

$$S = \frac{E - A - \mu_0 MH}{T} = \frac{\dfrac{3}{2}NkT + NkT\left\{1 + \ln\left[\left(\dfrac{2\pi mkT}{h^2}\right)^{\frac{3}{2}}\dfrac{V}{N}\right]\right\} - \mu_0 MH}{T} \qquad (43)$$

and by rearranging of (43) we can write

$$S = Nk\left\{\frac{5}{2} + \ln\frac{V}{N} + \frac{3}{2}\ln T + \frac{3}{2}\ln\frac{2\pi mk}{h^2}\right\} - \frac{\mu_0 MH}{T} \qquad (44)$$

Assuming approximation of ideal gas, the eq. (44) allows obtain the dependence of the temperature with the pressure of saturated vapor, then replacing (39) in (44)

$$S_{vapor} = Nk\left\{\frac{5}{2} + \ln\frac{kT}{P} + \frac{3}{2}\ln T + \frac{3}{2}\ln\frac{2\pi nk}{h^2}\right\} - \frac{\mu_0 MH}{T} \tag{45}$$

and for $P$ we obtain

$$\ln P = \frac{5}{2} + \frac{5}{2}\ln T + \ln\left[\left(\frac{2\pi m}{h^2}\right)^{\frac{3}{2}} k^{\frac{5}{2}}\right] - \frac{\mu_0 MH}{NkT} - \frac{S_{vapor}}{Nk} \tag{46}$$

The equation (46) will now be applied the case of saturated vapor in equilibrium with its condensed phase which in this study is solid phase. According Mandl [7], if $L$ is the molar enthalpy of sublimation at temperature $T$, $S_{vapor}$ and $S_{solid}$ are the molar entropies of vapor and solid respectively we have

$$S_{vapor} - S_{solid} = \frac{L}{T} \tag{47}$$

and the molar entropy of vapor phase $S_{vapor}$ is rewritten as

$$S_{vapor} = R\left\{\frac{5}{2} + \frac{5}{2}\ln T - \ln P + \ln\left[\left(\frac{2\pi m}{h^2}\right)^{\frac{3}{2}} k^{\frac{5}{2}}\right]\right\} - \frac{\mu_0 MH}{T} \tag{48}$$

Where $R = Nk$ and $N$ is Avogadro's Number.

Substituting (33) and (48) in equation (47) we obtain

$$R\left\{\frac{5}{2} + \frac{5}{2}\ln T - \ln P + \ln\left[\left(\frac{2\pi m}{h^2}\right)^{\frac{3}{2}} k^{\frac{5}{2}}\right]\right\} - \frac{\mu_0 MH}{T} - R[3\ln T - 3\ln\theta + 3 + \ln 2] = \frac{L}{T} \tag{49}$$

and $\ln P$ is given by

$$\ln P = -\frac{1}{2} - \frac{1}{2}\ln T + 3\ln\theta + \ln\left[\left(\frac{2\pi m}{h^2}\right)^{\frac{3}{2}} k^{\frac{5}{2}}\right] - \frac{\mu_0 MH}{RT} - \ln 2 - \frac{L}{RT}$$

(50)

Then the expression for the vapor pressure of paramagnetic solid can be written as

$$P = \frac{k^{\frac{5}{2}}(2\pi m)^{\frac{3}{2}}\theta^3 e^{-\frac{\mu_0 MH}{RT}}}{2h^3 e^{\frac{1}{2}}} \frac{1}{\sqrt{T}} e^{-\frac{L}{RT}}$$

(51)

Rearranging (51) and for the eq. (2), we have for $P$

$$P = \frac{k^{\frac{5}{2}}(\pi m)^{\frac{3}{2}}\theta^3 e^{-\frac{\mu_0 \chi H^2}{RT}}}{h^3 e^{\frac{1}{2}}} \sqrt{\frac{2}{T}} e^{-\frac{L}{RT}}$$

(52)

## 5. Conclusions

In this paper has been deduced an expression for the vapor pressure of a paramagnetic solid under conditions of high temperatures and weak field, which is a function of the magnetic susceptibility of solid and the external magnetic field applied. It is considered that vapor behaves as an ideal monoatomic gas and the entropy of the vibrations of crystalline solid was calculated in the limit of high temperature using the Einstein's model.

The statistical mechanics can enrich the courses of thermodynamics, which contributes to a better compression of the thermal phenomena. The thermodynamic equations deduced for the paramagnetic materials from the postulates of the statistical mechanics are tractable mathematically and offer a wide explanation of many physical systems of interest.